# Energy Content of Quantum Systems and the Alleged Collapse of the Wavefunction

Peter J. Riggs
Australian National University, Canberra ACT 0200, Australia
peter.riggs@anu.edu.au

**Abstract**

It is shown that within a quantum system, the wave field has a (potential) energy content that can be exchanged with quantum particles. Energy conservation in quantum systems holds if potential energy is correctly taken to be a field attribute. From this perspective, a transfer of energy occurs on measurement from the wave field to a quantum particle and this provides a physical explanation of what is commonly referred to as the 'collapse of the wavefunction'.

## 1. Introduction

Energy is fundamental to all physical processes. The role of energy in quantum systems, as distinct from its measurement, has been a somewhat neglected topic in quantum physics. A full account of the dynamics of a quantum system is only possible when all of a system's energy is taken into account. This needs to include the energy stored in a quantum system's wave field (matter wave) as it has become clear that a wave field is a real component of any quantum system [1]. Substantial evidence for the objective existence of wave fields has been steadily mounting, particularly over the last two decades [2]. It is not well known that a full account of the dynamics of quantum systems is possible within the deBroglie-Bohm Causal Theory of Quantum Mechanics, where a quantum system consists of particles embedded in a physical wave field [3]. It will be shown below that energy conservation holds in a quantum system and that energy changes occur within such a system when a measurement is made.

In deBroglie-Bohm Theory, the wavefunction $\Psi$ of a quantum system is a mathematical description of an objectively existing wave field. The wavefunction is expressed in polar form, i.e. $\Psi = R e^{iS/\hbar}$, where the wave field amplitude R and phase factor S are real-valued functions of the space and time coordinates. If the polar form is substituted into the single particle, time-dependent Schrödinger equation, then two coupled, real differential equations result. The relevant equation for current purposes is called the quantum Hamilton-Jacobi equation. In a one-particle quantum system, this equation is:

$$-\frac{\partial S}{\partial t} = \frac{(\nabla S)^2}{2m} + V - \frac{\hbar^2}{2m}\left(\frac{\nabla^2 R}{R}\right) \quad \ldots (1)$$

where *m* is the inertial mass of the quantum particle, V is an external (classical) potential, and other symbols have their usual meanings. Equation (1) differs from the classical Hamilton-Jacobi equation by having an extra term known as the quantum potential, denoted Q, i.e.

$$Q = (-\hbar^2/2m)(\nabla^2 R/R) \ldots (2)$$

Most of the differences between classical and quantum physics in deBroglie-Bohm Theory are accountable by the presence of the quantum potential. Analogous to the situation in classical Hamilton-Jacobi Theory, the momentum **p** of a (spinless) quantum particle is given by:

$$\mathbf{p} = \nabla S \ldots (3)$$

which ensures agreement with the empirical predictions of Orthodox Quantum Theory [4]. The particle's kinetic energy (T) can be seen from equation (3) to be the term $[(\nabla S)^2/2m]$ in equation (1). The particle's total rate of change of momentum with respect to time also can be shown to be [5]:

$$(d\mathbf{p}/dt) = -\nabla(V + Q) \ldots (4)$$

which reduces to $(-\nabla Q)$ in the classically-free case (i.e. when V = 0).

The relevant aspects of quantum systems will now be stated without further justification. It would require too much space to present the arguments in favour of these aspects and therefore the reader is referred to the cited references.

- Potential energy is a field attribute

Potential energy is conventionally defined as the potential energy of a particle or object. This usual definition is harmless in most physical contexts where it acts as a 'shortcut' to the actual location of potential energy in the field [6]. However, from a physical perspective, the field rather than the particle should be credited with potential energy [7]. A potential energy term should strictly be understood as a potential energy function that represents an amount of energy contained in a field and which is available to a particle situated within the field [8].

- Particles and field are intrinsic parts of a single system

Although quantum particles and their wave field may be dealt with separately for some analytical purposes, they are intrinsic parts of a single quantum system, i.e. the particles are not an optional addition to the field [9]. Treating quantum particles and wave field *exclusively* as separate but interacting entities will result in some incorrect conclusions being drawn.

- The wave field possesses energy-momentum

Since the wave field is a real field, it will have properties similar to other physical fields including having a finite spatial distribution and an energy density [10]. This energy density may be quite large, for the wave field can possess macroscopic orders of energy even when its wave-length is small [11]. The presence of energy-momentum in the wave field allows the field to affect the motion of quantum particles.

- Quantum Potential

The quantum potential Q (as defined by equation (2)) is the potential energy function of the wave field and represents that portion of the wave field's potential energy that is available to the particle at its specific position in the wave field [12].

- The wave field is not a mediated field

The wave field is not a mediated (or sourced) field unlike say, the electromagnetic field. A non-mediated field does not have a source term (such as electric charge) in its field equations, i.e. does not have a term not containing the field itself [13]. Therefore there is no 'quantum charge' and the wave field is not radiated [14].

## 2. Energy Content and Energy Conservation

The above aspects of quantum systems will now be applied in order to address energy issues in quantum systems. Where appropriate, examples will be presented utilising a Gaussian wave packet description. A Gaussian wave packet provides an excellent representation of the wave field since the field will be of finite extent and initially localised about the particle.

When expressed in terms of the real-valued functions R and S, the Hamiltonian density associated with the one-particle, time-dependent Schrödinger equation in the classically-free case is given by [15]:

$$\mathcal{H} = R^2(\nabla S)^2/2m + (\hbar^2/2m)(\nabla R)^2 \quad \ldots (5)$$

In the non-relativistic context, this is the total energy density. Let H be the quantity defined by:

$$H = \iiint_{-\infty}^{\infty} \mathcal{H} \, d^3\mathbf{x} \quad \ldots (6)$$

This integration yields a constant value for H when V = 0 [16]. H is the total energy of the quantum system (not just the energy content of the wave field) as the particle receives energy from the wave field and therefore the field energy cannot itself be constant [17]. Since particle and wave field are intrinsic parts of a single quantum system, the *total* energy content will be conserved in any classically-free, isolated system [18].

The quantity H being the total energy of an isolated, classically-free quantum system is particularly apparent in the example of a Gaussian wave packet. The wavefunction for a Gaussian packet at time t (> 0) is given by:

$$\Psi(\mathbf{x}, t) = (2\pi s_t^2)^{-3/4} \exp\{i\mathbf{k} \cdot (\mathbf{x} - \tfrac{1}{2}\mathbf{u}t) - (\mathbf{x} - \mathbf{u}t)^2/4\sigma_o s_t\} \quad \ldots (7)$$

where **u** is the initial group velocity, $\sigma_o$ is the initial root-mean-square (RMS) width of the packet in each coordinate direction, with $s_t = \sigma_o(1 + i\hbar t/2m\sigma_o^2)$, and other symbols have their usual meanings [19]. As before, if we let $\Psi = R e^{iS/\hbar}$, then we have from equation (7):

$$R = (2\pi\sigma^2)^{-3/4} \exp[-(\mathbf{x} - \mathbf{u}t)^2/4\sigma^2] \quad \ldots (8)$$

$$(\nabla R) = -(2\pi\sigma^2)^{-3/4}[(\mathbf{x} - \mathbf{u}t)/2\sigma^2]\exp[-(\mathbf{x}-\mathbf{u}t)^2/4\sigma^2]$$

$$= -R[(\mathbf{x} - \mathbf{u}t)/2\sigma^2] \quad \ldots (9)$$

$$(\nabla S) = m\mathbf{u} + (\hbar^2 t/4m\sigma_o^2\sigma^2)(\mathbf{x} - \mathbf{u}t) \quad \ldots (10)$$

$$(dS/dt) = \tfrac{1}{2}m|\mathbf{u}|^2 - (3\hbar^2 t/4m\sigma^2) + (\hbar^2 t/4m\sigma_o^2\sigma^2)[\mathbf{u}\cdot(\mathbf{x} - \mathbf{u}t)]$$

$$+ (\hbar^2/8m\sigma_o^2\sigma^2)(\mathbf{x} - \mathbf{u}t)^2 \quad \ldots (11)$$

and the particle's kinetic energy is:

$$T = (\nabla S)^2/2m$$

$$= \tfrac{1}{2}m|\mathbf{u}|^2 + (\hbar^2 t/4m\sigma_o^2\sigma^2)[\mathbf{u}\cdot(\mathbf{x}-\mathbf{u}t)] + (\hbar^4 t^2/32m^3\sigma_o^4\sigma^4)(\mathbf{x}-\mathbf{u}t)^2 \quad \ldots (12)$$

where $\sigma = |s_t| = \sigma_o[1+(\hbar t/2m\sigma_o^2)^2]^{1/2}$ is the RMS width of the packet at time t [20]. Using equations (5), (6), (8), (9) and (12), we find that the total energy of this (Gaussian) quantum system is:

$$H = (2\pi\sigma^2)^{-3/2}\iiint_{-\infty}^{\infty}\{\tfrac{1}{2}m|\mathbf{u}|^2 + (\hbar^2 t/4m\sigma_o^2\sigma^2)[\mathbf{u}\cdot(\mathbf{x}-\mathbf{u}t)]$$

$$+ [(\hbar^4 t^2/32m^3\sigma_o^4\sigma^4) + (\hbar^2/8m\sigma^4)](\mathbf{x}-\mathbf{u}t)^2\}\exp[-(\mathbf{x}-\mathbf{u}t)^2/2\sigma^2]\,d^3\mathbf{x}$$

$$= \tfrac{1}{2}m|\mathbf{u}|^2 + (3\hbar^4 t^2/32m^3\sigma_o^4\sigma^2) + (3\hbar^2/8m\sigma^2)$$

$$= \tfrac{1}{2}m|\mathbf{u}|^2 + (3\hbar^2/8m\sigma^2)[(\hbar^2 t^2/4m^2\sigma_o^4) + 1]$$

Using $(\sigma^2/\sigma_o^2) = [1 + (\hbar^2 t^2/4m^2\sigma_o^4)]$ we get:

$$H = \tfrac{1}{2}m|\mathbf{u}|^2 + (3\hbar^2/8m\sigma_o^2) \quad \ldots (13)$$

The first term on the right-hand side of equation (13) is obviously the particle's initial kinetic energy. The second term can be seen to be the initial field energy as it contains the initial RMS width of the wave packet and does not involve any velocities or time-dependent quantities. This confirms H as the total (constant) energy of the whole quantum system [21].

It can now be seen that the energy content of a (classically-free) wave field is given by (H – T). From equations (12) and (13), the total energy for a classically-free Gaussian wave field is:

$$(H - T) = (3\hbar^2/8m\sigma_o^2) - (\hbar^2 t/4m\sigma_o^2\sigma^2)[\mathbf{u}\cdot(\mathbf{x} - \mathbf{u}t)]$$

$$- (\hbar^4 t^2/32m^3\sigma_o^4\sigma^4)(\mathbf{x} - \mathbf{u}t)^2$$

$$= (3\hbar^2/8m\sigma_o^2) - (\hbar^2 t/4m^2\sigma_o^2\sigma^2)[m\mathbf{u} + (\hbar^2 t/8m\sigma_o^2\sigma^2)(\mathbf{x} - \mathbf{u}t)]\cdot(\mathbf{x} - \mathbf{u}t)$$

$$= (3\hbar^2/8m\sigma_o^2) - (\hbar^2 t/4m^2\sigma_o^2\sigma^2)[m\mathbf{u} + (\hbar^2 t/4m\sigma_o^2\sigma^2)(\mathbf{x} - \mathbf{u}t)]\cdot(\mathbf{x} - \mathbf{u}t)$$

$$+ (\hbar^4 t^2/32m^3\sigma_o^4\sigma^4)(\mathbf{x} - \mathbf{u}t)^2$$

$$= (3\hbar^2/8m\sigma_o^2) + (\hbar^2 t/2m\sigma_o^2)[(\nabla S)/m]\cdot[(\nabla R)/R]$$

$$+ (1/2m)(\hbar^2 t/2m\sigma_o^2)^2[(\nabla R)/R]^2$$

where equations (8), (9) and (10) have been employed.

    We can express (H − T) more generally in terms of the wave field's amplitude R, its derivatives and the derivatives of S. We shall make use of the identity:

$$\nabla\cdot\left(\frac{\nabla R}{R}\right) = \frac{R(\nabla^2 R) - (\nabla R)^2}{R^2} = -\frac{3}{2\sigma^2}$$

so that

$$\frac{\hbar^2 t}{2m\sigma_o^2} = -(\nabla^2 S)\bigg/\left[\nabla\cdot\left(\frac{\nabla R}{R}\right)\right] = \frac{R^2(\nabla^2 S)}{(\nabla R)^2 - R(\nabla^2 R)}$$

where $(\nabla^2 S) = (3\hbar^2 t/4m\sigma_o^2\sigma^2)$ from equation (10). Now since $\nabla^2(dS/dt) = (3\hbar^2/4m\sigma_o^2\sigma^2)$ from equation (11), we find that:

$$\frac{3\hbar^2}{8m\sigma_o^2} = \left(\frac{-3}{4}\right)\nabla^2\left(\frac{dS}{dt}\right)\bigg/\left[\nabla\cdot\left(\frac{\nabla R}{R}\right)\right] = \left(\frac{3}{4}\right)\frac{R^2\nabla^2(dS/dt)}{(\nabla R)^2 - R(\nabla^2 R)}$$

Hence (H − T) =

$$\left(\frac{3}{4}\right)\frac{R^2\nabla^2(dS/dt)}{(\nabla R)^2 - R(\nabla^2 R)} + \frac{R^2(\nabla^2 S)}{(\nabla R)^2 - R(\nabla^2 R)}\left(\frac{\nabla S}{m}\right)\cdot\left(\frac{\nabla R}{R}\right)$$

$$+ \left(\frac{1}{2m}\right)\left[\frac{R^2(\nabla^2 S)}{(\nabla R)^2 - R(\nabla^2 R)}\right]^2\left(\frac{\nabla R}{R}\right)^2 \quad \ldots (14)$$

Equation (14) shows an explicit time dependence as would be expected, i.e. the wave field's energy content varies over time. Further, equation (14) shows that the energy content of the wave field is independent of the field's intensity (where intensity is proportional to the square of the wave's amplitude R), as is also the quantum potential. This can be seen because multiplication of the amplitude R by a constant does not change the value of (H − T) or of Q [22]. This indicates that, unlike a classical field, the wave field's form has greater physical significance than its amplitude.

It is claimed in the literature that energy is not conserved for a quantum system as a whole, i.e. wave field and particle together, in deBroglie-Bohm Theory [23]. There are two premises used in reaching this conclusion. The first is the lack of a classical reaction on the wave field, for although the wave field acts on the particle, the quantum particle does not affect the size or shape of the field. However, non-mediated fields cannot display classical reactions due to the absence of source terms. This lack of a classical reaction need not disallow the possibility of energy conservation for a quantum system as a whole. The second premise used is that potential energy is a particle property. Defining potential energy as a field attribute is not only more satisfactory physically but is essential when dealing with non-mediated fields. Energy conservation can be shown to hold in classically-free quantum systems when potential energy is correctly defined to be a field characteristic.

In order to see this more clearly, let the energy content of the wave field (other than that given by the quantum potential) in an isolated, classically-free, non-stationary state be U, then U = H − T − Q. The total time rate of change of U is then given by:

$$\frac{d\text{U}}{dt} = \frac{d\text{H}}{dt} - \frac{d\text{T}}{dt} - \frac{d\text{Q}}{dt} \quad \ldots (15)$$

Now

$$\frac{d\text{Q}}{dt} = \sum_{i=1}^{3} \frac{\partial \text{Q}}{\partial x^i} \frac{dx^i}{dt} + \frac{\partial \text{Q}}{\partial \text{t}} = (\nabla \text{Q}) \cdot \left(\frac{\nabla \text{S}}{m}\right) + \frac{\partial \text{Q}}{\partial \text{t}}$$

$$= - \frac{d\text{T}}{dt} + \frac{\partial \text{Q}}{\partial t} \quad \ldots (16)$$

where $(d\text{T}/dt) = [- (\nabla \text{Q}) \cdot (\nabla \text{S}/m)]$ as can be seen by differentiating $[(\nabla \text{S})^2/2m]$ and using equations (3) and (4). When $(\partial \text{Q}/\partial \text{t}) = 0$, the energy available to the particle (T + Q) is constant and any change in kinetic energy is exactly balanced by a change in the quantum potential. Thus changes in the particle's kinetic energy is a straight-forward energy conversion process.

The situation is more complicated when $(\partial \text{Q}/\partial \text{t}) \neq 0$. Equations (15) and (16) together give the result:

$$(d\text{U}/dt) = - (\partial \text{Q}/\partial t) \quad \ldots (17)$$

since $(d\text{H}/dt) = 0$ for a classically-free system. Equation (17) shows that $(\partial \text{Q}/\partial t)$ gives the change of the quantum potential due to changes in U, i.e. increases/ decreases in Q are due to decreases/ increases in the amount of energy stored in the wave field other than at the particle's location [24]. Changes in a quantum particle's kinetic energy are now more generally explained by an energy conversion process where the quantum potential is the 'channel' for energy to and from the wave field to the particle and back again. Energy conservation holds in an isolated, classically-free, single particle quantum system by taking account of such transfers.

The total time rate of change of U may be expressed in terms of Q, amplitude R and its derivatives. From equations (2) and (17), we find:

$$\frac{d\mathrm{U}}{dt} = -\left(\frac{\partial \mathrm{Q}}{\partial t}\right) = \frac{\hbar^2}{2m}\frac{\partial}{\partial t}\left(\frac{\nabla^2 \mathrm{R}}{\mathrm{R}}\right) = \frac{\hbar^2}{2m\mathrm{R}}\nabla^2\left(\frac{\partial \mathrm{R}}{\partial t}\right) + \frac{\mathrm{Q}}{\mathrm{R}}\left(\frac{\partial \mathrm{R}}{\partial t}\right) \quad \dots \quad (18)$$

The term ($\partial$R/$\partial$t) gives the rate of change of R explicitly due to changes over time in the shape of the wave field. Equation (18) shows that ($\partial$Q/$\partial$t) ≠ 0 over the time interval of a change in the shape of the wave field. It can be seen from this that the more pronounced the change in shape is, the greater will be the amount of energy exchanged between particle and wave field.

## 3. Alleged Wavefunction Collapse

What is commonly called 'measurement' in Orthodox Quantum Theory is only one type of interaction between different physical systems. In the single-particle deBroglie-Bohm account of the measurement process, the wavefunction is split into non-overlapping wave packets that are independent of each other. One packet will contain the particle and will affect its motion whilst the other (empty) wave packets will not. There is no 'collapse of the wavefunction'. The interaction with a device designed to measure an 'observable' of a quantum system is described in deBroglie-Bohm Theory by means of a multi-dimensional configuration space as the interaction involves a many-particle system (viz. the measurement apparatus) [25]. Although the mathematical description is in terms of such a configuration space, the actual measurement interaction occurs, not in a multi-dimensional configuration space nor in a Hilbert space, but in physical (three-dimensional) space. Therefore, we should be able to give an account of what happens to particle and field in physical space during the conduct of a measurement.

Measurement processes generally introduce disturbances to a quantum system. In order to visualise what occurs during measurement, consider a quantum particle trapped in a cubical box with zero classical potential inside. In this situation, the wave field will be a standing wave with the quantum particle at rest inside the box [26]. If one of the walls of the box is then removed as part of a measurement process, Orthodox Quantum Theory postulates that there will be an instantaneous 'collapse of the wavefunction' as this will only reveal a quantum particle with those specific properties that are being measured. In deBroglie-Bohm Theory, there is a physical change caused to the wave field on measurement with the following results: (i) the wavefunction (which describes the wave field) is correspondingly altered; and (ii) particle attributes which depend on the wave field (e.g. momentum) will generally be changed as a consequence. When the wall is removed, the particle and its wave field will no longer be constrained to remain in the box. This quantum system will evolve from a stationary state and its wave field will change in form from a standing wave to a travelling wave as it propagates out of the box. There is also a change in amplitude that accompanies the change in the shape of the wave field immediately after the wall of the box is removed. From equation (18), we have:

$$\frac{d\mathrm{U}}{d\mathrm{t}} = \frac{\hbar^2}{2m\mathrm{R}} \nabla^2 \left(\frac{\partial \mathrm{R}}{\partial \mathrm{t}}\right) + \frac{\mathrm{Q}}{\mathrm{R}}\left(\frac{\partial \mathrm{R}}{\partial \mathrm{t}}\right)$$

This indicates that, during this transition process, energy will transfer from the wave field to the particle which will then appear as its kinetic energy. The more pronounced the change in shape is, as indicated by non-zero values of the term $(\partial \mathrm{R}/\partial \mathrm{t})$, the greater will be the amount of energy gained by the particle from the wave field [27].

Upon emerging from the open part of the box, a physically reasonable approximation for the form of the wave field is represented by the initial, normalised Gaussian wavefunction (i.e. equation (7) with time t set to zero):

$$\Psi_o = (2\pi\sigma_o^2)^{-3/4} \exp\{i\mathbf{k} \cdot \mathbf{x} - (|\mathbf{x}|^2/4\sigma_o^2)\}$$

The initial quantum potential corresponding to $\Psi_o$ is [28]:

$$Q_o = (\hbar^2/4m\sigma_o^2)\{3 - (|\mathbf{x}_o|^2/2\sigma_o^2)\}$$

Suppose the cubical box is of side length L and take one corner of the box as the origin of a rectangular Cartesian coordinate system, then $\sigma_o = (L/2)$ and the particle's position will be at the end of the box, i.e. $|\mathbf{x}_o| = L$. This gives a value of: $Q_o = (\hbar^2/4mL^2)$. A calculation of the value of the quantum potential when the wave field is in a stationary state inside the closed box shows its value to be $(3\pi^2\hbar^2/2mL^2)$ [29]. The difference between these two values of the quantum potential shows a decrease of slightly less than $(15\hbar^2/mL^2)$ which becomes the kinetic energy of the quantum particle. Once clear of the obstruction (i.e. the open end of the box) the wave field will tend to the form of a travelling plane wave of constant amplitude (i.e. a free particle wave field). Note that this transfer of energy to the particle is not instantaneous and does not violate the Special Theory of Relativity [30].

Since a measurement in these circumstances will only reveal a quantum particle with definite properties, Orthodox Quantum Theory has to postulate 'wavefunction collapse' to formally achieve the result that is found by an actual measurement. The alleged 'collapse of the wavefunction' can be seen here merely to be a *surrogate* for the physical situation where the wave field loses energy to a quantum particle.

## 4. Conclusions

The wave field of a quantum system is a physical field that stores energy. The quantum potential is a potential energy function which represents the amount of energy contained in the wave field that is available to particles within the field. Energy conservation in quantum systems can be shown to apply if potential energy is correctly held to be a field attribute and the role of the quantum potential in 'channelling' energy is taken into account. The alleged 'collapse of the wavefunction' on measurement postulated in Orthodox Quantum Theory does not occur. Instead there is a transfer of energy from the wave field to the quantum particle during the measurement process.

# References

1. Riggs, P.J. *Quantum Causality: Conceptual Issues in the Causal Theory of Quantum Mechanics* (Dordrecht: Springer, 2009), chapter 4.

2. Kozuma, M. et al. *Science* **286**, 2309 (1999); Ketterle, W. *MIT Physics Annual 2001* (Cambridge, MA.: Massachusetts Institute of Technology, 2001); Bongs, K. and Sengstock, K. *Reports on Progress in Physics* **67**, 907 (2004).

3. Holland, P.R. *The Quantum Theory of Motion: An Account of the deBroglie-Bohm Causal Interpretation of Quantum Mechanics* (Cambridge: Cambridge University Press, 1993) chapter 3; Shifren, L. et al. *Technical Proceedings of the 2001 International Conference on Modeling and Simulation of Microsystems* (Nano Science and Technology Institute, 2001); Garashchuk, S. and Rassolov, V.A. *Journal of Physical Chemistry A* **111**, 10251 (2007); Riggs, P.J. *Quantum Causality*, 95-96.

4. Bohm, D. *Physical Review* **85**, 170 (1952).

5. Riggs, P.J. *Quantum Causality*, 58-59.

6. Rindler, W. *Relativity: Special, General, and Cosmological* (Oxford: Oxford University Press, 2006) 113; Riggs, P.J. *Quantum Causality*, 91-94.

7. Rindler, W. *Introduction to Special Relativity* (Oxford: Pergamon, 1982) 132.

8. Riggs, P.J. *Erkenntnis* **68**, 21 (2008).

9. Riggs, P.J. *Quantum Causality*, 52, 104.

10. Holland, P.R. *The Quantum Theory of Motion*, 84.

11. Bohm, D. in: *Scientific Papers Presented to Max Born (*Edinburgh and London: Oliver & Boyd, 1953) 14.

12. Riggs, P.J. *Quantum Causality*, 114.

13. Doughty, N.A. *Lagrangian Interaction: An Introduction to Relativistic Symmetry in Electrodynamics and Gravitation* (Sydney: Addison-Wesley, 1990) 139.

14. Holland, P.R. *The Quantum Theory of Motion*, 83.

15. Ibid., 115.

16. Ibid., 116.

17. Riggs, P.J. *Quantum Causality*, 116.

18. Ibid.

19. Holland, P.R. *The Quantum Theory of Motion*, 159.

20. Riggs, P.J. *Journal of Physics A* **32**, 3071 (1999).

21. Riggs, P.J. *Quantum Causality*, 121.

22. Ibid., 128.

23. Holland, P.R. *The Quantum Theory of Motion*, 120.

24. Riggs, P.J. *Quantum Causality*, 117.

25. Holland, P.R. *The Quantum Theory of Motion*, 339-342; Riggs, P.J. *Quantum Causality*, 62-65.
26. Riggs, P.J. *Quantum Causality*, 111.
27. Ibid., 134.
28. Ibid.
29. Ibid., 135.
30. Ibid., 133-134.